\documentclass{epl}
\usepackage{graphicx}
\title{Chaotic and power law states in the Portevin-Le Chatelier effect}
  \author{M.S. Bharathi\inst{1} \and G. Ananthakrishna\inst{1,2}}
  \institute{
    \inst{1} Materials Research Centre, Indian Institute of Science,
 Bangalore-560012, India\\
    \inst{2} Centre for Condensed Matter Theory, Indian Institute of
 Science, Bangalore-560012, India
  }

\begin{document}

  \maketitle

  \begin{abstract}
  Recent studies on the Portevin - Le Chatelier effect report an intriguing
crossover phenomenon from a low dimensional chaotic to an infinite
dimensional scale invariant power law regime in experiments on CuAl single
crystals and AlMg polycrystals, as a function of strain rate. We devise a
fully dynamical model which reproduces these results. At low and medium
strain rates,   the model is chaotic with the structure of the attractor resembling the
reconstructed experimental attractor.  At high strain rates, power law
statistics   for the magnitudes and durations  of the  stress drops emerge as in
 experiments   and concomitantly, the  largest Lyapunov exponent is zero.

  \end{abstract}

  The Portevin-Le Chatelier (PLC) effect, discovered at
  the turn of the last century \cite{PLC},  is a striking
  example where collective behaviour of defects leads
  to complex spatio-temporal patterns \cite{Set,Kubin01,Cottrell}.
  The PLC effect manifests itself as a series of serrations on the
stress-strain curves when samples   of dilute alloys are deformed under constant strain rate,
$\dot{\epsilon}_a$ (actually constant pulling speed).
  The effect is observed only in a window of strain rates and temperatures.
  Each stress drop is associated with the formation and often the
propagation of a dislocation band.    In polycrystals, at low $\dot{\epsilon}_a$,
 the randomly nucleated type C bands with
  large stress drop amplitudes are seen. At intermediate strain rates, one
finds the spatially   correlated 'hopping'  type B bands moving in a relay race manner with
 smaller   stress drop amplitudes. At high strain rates, propagating type A bands
with small amplitudes are observed. (In single crystals such a clear
classification does not exist.)  These different types of PLC bands are
believed to represent distinct correlated states of dislocations in the bands. It
is this rich spatio-temporal dynamics that has recently attracted the attention of
physicists as well \cite{Danna,Mertens}. Indeed, the PLC effect is a good example of
slow-fast dynamics commonly found in many stick-slip systems such as
frictional sliding \cite{Pers}, fault dynamics \cite{Carlson} and
peeling of an   adhesive tape \cite{Maug}.

Recent efforts have shown that surprisingly large
 body of information about the nature of dynamical correlations is  hidden
  in the stress-strain curves \cite{Anan95,Noro97,Anan99,Bhar01}. More
 recently, an {\it intriguing crossover phenomenon} from a
  chaotic regime occurring at low and medium strain
  rates to a power law regime at high strain rates has been
  detected in experiments on the PLC  effect
  in Cu-10\% Al single crystals \cite{Anan99} and Al-Mg polycrystals
\cite{Bhar01}.   This suggests that the {\it crossover itself is insensitive to the
microstructure.}    The purpose of this paper is to extend a model for the PLC effect
introduced earlier \cite{Anan82} to explain this crossover phenomenon.

  This crossover phenomenon is of interest in the larger context of dynamical
  systems as this is a rare example of a transition between
  two dynamically distinct states. Chaotic systems
  are characterised by the self similarity of the strange attractor
  and sensitivity to initial conditions quantified by fractal dimension and
  the existence of a positive Lyapunov exponent, respectively.
  In contrast, a class of spatially extended dissipative systems often exhibit a tendency to evolve to a marginally
stable state,  characterised by power law statistics for the events, under the action
  of slow external drive with no parameter tuning. (It must be emphasized
that power laws  can  arise due to other mechanisms also \cite{Sornette}.)
  Such a state is termed as  self-organised criticality (SOC) by Bak {\it et al} \cite{Bak}.
  Unlike the low dimensional nature of the chaotic attractor, SOC state
  is an infinite dimensional state.  Large number of physical systems are known to exhibit SOC
  type features \cite{Jensen,Bak96}.

  There are numerous models and experiments where {\it either of
  these dynamical regimes} have been detected \cite{Abar,Jensen,Bak96}.
  To the best of our knowledge ref. \cite{Anan99}  is one of the two
  instances known where both these states are observed in one and the same
system.  The other example is in hydrodynamics where chaos is
  observed at low Rayleigh number and power law scaling regime, known as
hard turbulence,  is seen at high values \cite{Lib}. Moreover, both the hard turbulence
  in convection and the power law regime in the PLC effect are observed
   at high drive  parameter values in contrast  to most SOC
  systems \cite{Jensen,Bak96}. Thus, power laws in the PLC effect are
 closer to the turbulent  regime in convection than other SOC systems.

  The microscopic origin of the PLC effect is due to the interaction of mobile
  dislocations  with diffusing solute atoms and is referred to
  as dynamic strain aging (DSA), first suggested by Cottrell
\cite{Cottrell} and later improved by others.
  (See for instance ref. \cite{Set,Kubin01}). At low strain rates
  (or high temperatures)  the average velocity of dislocations is low,
  there is sufficient time for solute atoms to diffuse to dislocations and
pin them (usually called as aging).
  Thus, longer the dislocations are arrested, larger will be the stress
required to unpin them.  When these dislocations are unpinned, they move at large speeds till they
are pinned again.  At high strain rates (or low temperatures), the time available for solute
atoms to diffuse to the  dislocations decreases and hence the stress required to unpin them
decreases. Thus, in a range of  strain rates and temperatures where these two time scales are typically
of the same order of  magnitude, the PLC  instability manifests.
  The competition between the slow rate of pinning and sudden unpinning of
the dislocations,  at the macroscopic level translates into a negative strain rate
sensitivity (SRS) of the flow stress as a  function of strain rate which  is the basic instability mechanism used in
  most phenomenological models \cite{Set,Kubin01}.

  The well separated time scales implied in the DSA is mimicked by the fast mobile, immobile and the intermediate
  'decorated' Cottrell type dislocations in the dynamical model due to
Ananthakrishna and  coworkers \cite{Anan82}.  The basic idea of the model is that qualitative
features of the PLC effect emerge from the nonlinear interaction of these few
dislocation populations, assumed to represent the collective degrees of freedom of the system.
  The rate equations for dislocation densities are constructed based on
known dislocation mechanisms.  In spite of the idealised nature of 
the model, it is successful in
explaining several generic features of the   PLC effect, such as the emergence of the negative
  SRS \cite{Anan82,Rajesh}, the existence of critical strain for the onset of
  the PLC instability, and the existence of a window of strain rates and
temperatures  for the occurrence of the PLC effect (see \cite{Anan82}).
  One prediction of the model is that there is a range of strain rates where
  the PLC effect is chaotic \cite{Anan83}, subsequently verified by
  analysing  experimental signals \cite{Anan95,Noro97,Anan99,Bhar01}. The 
model has
  been studied in detail by our group and others including an
  extension to the case of fatigue \cite{Bekele,Glazov,Zaiser}.

  Since the model is fully dynamical and it predicts chaos at intermediate
 strain  rates found in experiments,  it has the right   ingredients for studying
  this crossover.  Here, we outline the model in terms of scaled variables
in the notation of  Ref. \cite{Rajesh00}. In addition, we introduce a
  spatial coupling arising out of the cross-slip mechanism, used
earlier  by others \cite{Set,Kubin01}.  The model consists of densities of  mobile, immobile,
  and Cottrell's type dislocations denoted by
   $\rho_m(x,t)$, $\rho_{im}(x,t)$ and $\rho_c(x,t)$ respectively, in the
scaled form.  The  evolution equations  are:
  \begin{eqnarray}
  \frac{\partial{\rho_m}}{\partial t} & = & -b_0\rho_m^2 -\rho_m\rho_{im}+\rho_{im} - a \rho_m + \phi_{eff}^m\rho_m +
\frac{D}{\rho_{im}}\frac{\partial^2 (\phi_{eff}^m(x)\rho_m)}{\partial x^2},\\
  \frac{\partial{\rho_{im}}}{\partial t} & = & b_0(b_0\rho_m^2
-\rho_m\rho_{im} -\rho_{im}+a\rho_c), \\
  \frac{\partial{\rho_c}}{\partial t} & = & c(\rho_m-\rho_c).
  \end{eqnarray}
  \noindent
  The first term in eqn.(1) refers to the formation of locks and consequent
immobilisation of two  mobile dislocations, the  second term to the annihilation of a mobile
  dislocation with an immobile one, and   the  third term  to
  the remobilisation of the immobile dislocation due to stress or thermal
  activation. The fourth term represents the
  immobilisation of mobile dislocations due to solute atoms. Once a mobile
  dislocation starts acquiring  solute atoms we regard it as
  the Cottrell's type dislocation $ \rho_c$. As they progressively acquire
more solute atoms, they eventually stop, then
  they are considered as immobile dislocations $\rho_{im}$ (loss term
  in eqn. (3) and the gain term in eqn. (2)). Alternately, the 
aggregation 
of solute
atoms can be  regarded as the definition of $\rho_c$, ie.,
  $\rho_c = \int_{-\infty}^t dt^{\prime}\rho_m(t^{\prime} )
K(t-t^{\prime})$. For the sake of simplicity, we use a single time
  scale with $K(t) = e^{-ct} $ .  The convoluted nature of the
  integral physically implies that the mobile dislocations to which solute
atoms aggregate earlier  will be aged more than those which acquire solute atoms later ( see ref.
  \cite{Rajesh}). The fifth term  represents the rate of multiplication of
  dislocations due to cross-slip. This depends on the velocity of the
mobile dislocations taken to be $ V_m(\phi) = \phi_{eff}^m$, where $\phi_{eff} = (\phi - h \rho_{im}^{1/2})$
is the scaled   effective stress, $\phi$ the scaled stress, $m$ the velocity exponent
and $h$ a work hardening parameter. The
  last term is a spatial coupling term arising out of the nonlocal nature
of the cross-slip as argued  below. Actually, the nature of the spatial coupling in the PLC effect has
been a matter of debate \cite{Set,Kubin01}. Within the scope of the
  model, a natural  source of spatial coupling is the nonlocal nature of cross-slip as
dislocations generated at a point
  spread over to the neighbouring elements. (Compatibility stresses between
the slipped   and the unslipped regions and long range interactions are other
 possible sources of coupling \cite{Set,Kubin01}.)   Let $\Delta x$ be an elementary
 length. Then, the flux $\Phi(x)$ flowing from $x \pm \Delta x$
   and out of $x$ is given by $\Phi(x) + \frac{p}{2} \left[\Phi(x+\Delta x)
- 2 \Phi(x) + \Phi(x - \Delta x)\right]$ where   $\Phi(x) = \rho_m(x)V_m(x)$. Here $p$ is the probability of cross-slip
spreading into  neighbouring elements. Expanding $\Phi(x \pm \Delta x)$ and keeping the
leading terms, we get $\rho_m V_m +  \frac{p}{2}\frac{\partial^2(\rho_m V_m)}{\partial x^2} (\Delta x)^2.$
Further, cross-slip spreads  only into regions of minimum back stress. Noting that the back stress is
  usually taken to result from the immobile dislocation density ahead of it,
  we use $\Delta x^2 = <\Delta x^2> = \bar {r}^2 \rho_{ im}^{-1}$. Here $<\ldots>$ refers to the ensemble average
  and $\bar{r}^2$ is an elementary (dimensionless) length. Finally, $a$, $b_0$ and $c$
  are the scaled rate constants referring, respectively, to the
concentration of solute atoms slowing down the
  mobile dislocations, the thermal and athermal reactivation of immobile dislocations,
  and the rate at which solute atoms are gathering around the mobile
dislocations. We note that the order of
  magnitudes of the constants have been identified in ref. \cite{Anan82,Bekele}.
  These equations are coupled to the machine equation
  \begin{equation}
  \frac{d\phi(t)}{dt}= d [\dot{\epsilon}-\frac{1}{l}\int_0^l\rho_m(x,t)
  \phi_{eff}^m(x,t)dx],
  \end{equation}
 where $\dot\epsilon$ is the scaled applied strain rate,
  $d$ the scaled effective modulus of the machine and the sample,
  and $l$ the  dimensionless length of the sample.
  (We reserve $\dot {\epsilon}_a $ for the unscaled strain rate.)

  The PLC state is reached through a Hopf bifurcation.
  The domain of instability in  $\dot\epsilon$  is  $10 < \dot \epsilon < 2000$,
  and that in other parameters is  the same as in the original model,
  beyond which uniform steady states exist .
 Here, we use $a = 0.8,
  b_0 = 0.0005, c = 0.08, d =0.00006, m = 3.0, h = 0.07$ and $D=0.5$.
  (The parameter values used here are essentially the same used in several of
  our earlier calculations. \cite{Rajesh00}.)
  But  the results discussed below hold true for a wide range of other
 parameters  in instability domain including a range of values of $D$.

  These equations are discretised into $M$ equal parts of width  ${\Delta}l$
  and solved numerically for $\rho_m(j,t)$, $\rho_{im}(j,t), \rho_c(j,t)$,
$j= 1,2,...,M$,   and ${\phi}(t)$.  Due to the widely differing time scales, appropriate
care has been exercised  in the numerical solutions of these equations by using a variable step
fourth order   Runge-Kutta scheme with an accuracy of $10^{-4}$ for all the four
variables.  The spatial derivative of $\rho_m$ is approximated by its central
difference.  The initial values of dislocation densities are taken to be
  their steady state values with a Gaussian spread along the length of the
sample with a Gaussian spread along the length of the sample,
since the long term evolution of the system is essentially independent of
the initial values.  Now consider the boundary conditions. Since the sample is strained at the
grips, we choose   $\rho_{im}(j,t)$ at $j=1$ and $N$ to be three orders of magnitude more
than rest of the sample and  $\rho_m(j,t) =\rho_c(j,t)=0$ at $j =1$ and $N$, as $\rho_m$ and $\rho_c$
cannot evolve into the grips .

  We first recall the relevant experimental results on the crossover
  phenomenon and compare them with those from the
  model. Plots of two experimental stress-strain curves corresponding to
the chaotic  and SOC regimes    of applied strain rates are shown in Fig. 1.
  In ref. \cite{Anan99}, the chaotic nature of the stress-strain curve shown in
  Fig. 1a, was demonstrated by showing the existence of a finite
correlation dimension using the standard
  Grassberger-Procaccia algorithm \cite{Grass}, and the existence of
   a positive Lyapunov exponent \cite{Abar}. Both these methods involve
embedding the time series in a higher  dimensional space using time-delay technique \cite{Abar}.
  (In addition, surrogate data analysis was also carried out in
\cite{Anan99}.) The correlation dimension, $\nu$,
  of the experimental attractor was found to be 2.3.
  Then, the number of degrees of freedom required for the description
  of the dynamics of the system is  given by the minimum integer larger than $\nu + 1$ which is four in this
case, consistent with that used in the original model.
  The geometrical interpretation of these degrees of freedom is that
  it is the subspace to which the  trajectories are confined. This dimension can also be obtained by an
  alternate method, called the singular value decomposition \cite{King},
  which has an {\it additional advantage} of allowing the {\it
visualisation  of the strange attractor}. This method has been applied to the PLC time
series earlier \cite{Noro97}. In this method,
  the trajectory matrix is  constructed and the eigen values of the covariance matrix are calculated.
  For the time series in Fig. 1a, we find that the relative strength
  of the fourth eigen value drops more than two orders of magnitude
compared  to the first and changes very little beyond the fourth eigen value.
  Thus, we estimate the dimension of the experimental
  attractor to be four, consistent with that obtained from the correlation
dimension.   Using the first three principal directions of the subspace $C_i; i =1$ to
3,  we have reconstructed the experimental attractor in the space of
  specifically chosen directions $C_1 -C_2, C_3$ and $C_1$ to permit
comparison  with the model. This is shown in Fig. 2a for the
  experimental time series   at $\dot {\epsilon}_a = 1.7 \times 10^{-5}$ s $^{-1}$.
  This can be compared with the strange attractor obtained from the model
  in the space of $\rho_m,\rho_{im}$ and $\rho_c$ (at an
  arbitrary spatial location, here $j=50$ and $N=100$)
  shown in Fig. 2b for $\dot \epsilon = 120$ corresponding to
  the mid chaotic region (see below).  Note the   {\it similarity with the experimental attractor} particularly
  about the linear portion in the phase space (Fig. 2a)  identified
  with the loading direction in Fig. 1a.  Note that the identificaton of 
the loading direction is consistent with the absence of
growth of $\rho_m$. Since the existence of a positive
  Lyapunov exponent is a confirmatory test of chaos,    we have calculated both the largest Lyapunov exponent (LLE) and the
  spectrum of Lyapunov exponents \cite{Abar} directly from the model.
  Since the magnitude of the LLE and that of the largest Lyapunov exponent
calculated from the  spectrum agree, we have shown the average LLE (obtained over 15000 time
  steps after stabilisation) in Fig. 3a as a function of the strain rate 
for $N=100$.
  The LLE becomes positive around $\dot \epsilon = 35$ reaching a maximum
at  $\dot \epsilon =120$, and practically
  vanishing around 250. (Periodic states are observed in the interval $10
  <\dot \epsilon < 35$.) For $\dot \epsilon \ge 250$, the dispersion in the
  value of the LLE is  $ \sim 5 \times 10 ^ {-4}$ which is the same order as the mean.
  Thus, the LLE is taken to vanish beyond $\dot\epsilon = 250$ as is
expected of  the power law regime \cite{Obukh}.

  At high strain rates beyond $\dot\epsilon \sim 280$,
  the stress-time series obtained from the model shows no inherent
  scale in the magnitudes of the stress drops as in the
  case of experimental time series at high strain rates
\cite{Anan99,Bhar01}.    We have analysed the distributions for the stress drop magnitudes
$\Delta\phi$ and  their durations $\Delta t$.  The distribution of stress drop magnitudes,
  $D(\Delta \phi)$, shows a power law $D(\Delta \phi) \sim \Delta \phi ^{-
\alpha }$.  This is shown in Fig. 3b($\circ$) along with the experimental
  points ($\bullet$) corresponding to $\dot {\epsilon}_a = 8.3 \times 10^{-5} s^{-1}$.
  Clearly both experimental and theoretical points
  show a scaling behaviour with an exponent value $\alpha \approx 1.1$.
  The distribution of the durations of the drops $D(\Delta t) \sim \Delta t^{ -\beta}$
  also shows a power law with an exponent value $\beta \approx 1.3$.
  The conditional average of $\Delta \phi $ denoted by $<\Delta\phi>_c$ for
a given value of
  $\Delta t$ behaves as $<\Delta\phi>_c \sim {\Delta}t^{1/x}$ with $x \approx 0.65$.
  The exponent values satisfy the scaling relation $\alpha = x(\beta - 1)+1$ quite well.
  Since the basic cause of the stress-drops is the growth of
  mobile dislocation density during this period, one can look at the scaling
  behaviour of the total density in the sample at a given time using
  $\bar{\rho}_m(t)= \int \rho_m(x,t) dx $.
  Let $\Delta \bar{\rho}_m (t)$ denote the increase
  in $\bar {\rho}_m(t)$ occurring during the intervals of the stress drops.
  Then, one should expect that the statistics
  of $\Delta \bar{\rho}_m$ also to exhibit a power law, ie.,
  $D(\Delta \bar{\rho}_m) \sim \Delta \bar{\rho}_m^{-\gamma} $,
  with the same exponent value as $\alpha$.  
A plot of $ D(\Delta \bar{\rho}_m)$ for $N=300$ is shown in Fig. 3b 
($\diamondsuit$).
  The extent of the power law regime is nearly two orders with $\gamma \approx 1.1 $,
  same as $\alpha$.  The large bump at high values is due to the effect of
finite size of the system as in many models
   \cite{Carlson} and in particular, here it is due to high
levels of stress at the grips. Noting that dislocation bands cannot
propagate into regions of high stresses, it is clear that the edges cause
distortions in the otherwise smoothly propagating bands leading to large
changes in $\bar\rho_m(t)$. Typically, the influence of the edges are felt
by the band when it is 20 sites away. Increasing $N$ from 100 to 300,
increases the scaling regime by half a decade and the peak of the bump
reduces from 700 to 500, thus indicating the influence of the finite size
of the system.

  To sum up, the present model exhibits chaotic dynamics
  at low and medium strain rates, and SOC dynamics at high $\dot\epsilon$.
  These distinct dynamical states and the crossover clearly
  emerge due to the inclusion of nonlinearities
  in the form of basic dislocation mechanism and a spatial coupling.
  As for the nature of dynamics in different regions of strain rate,
  one can perhaps anticipate the emergence of the spatio-temporal chaotic
regime \cite{Hohen},  as the original model exhibits chaos. However, the emergence of SOC
dynamics  needs some explanation. Recently, we have shown that the upper limit
  of the PLC effect in the original model is a result of a reverse
  Hopf bifurcation at high strain rates \cite{Rajesh00}.
   This implies that the average amplitude of the stress drops decreases
  as a function of strain rate as in experiments.  This means that the average stress level
  (in time) is roughly constant with small fluctuations around the mean
implying that  $\dot\epsilon$ essentially balances the total plastic strain rate.
  The picture essentially remains unaltered when spatial coupling is
introduced.    Thus, in this regime, the state of the system is critically poised as in
any SOC system.  Recently, we have shown that the geometry
  of the slow manifold has a bent structure and the 'fold line' on the slow
manifold  corresponds to the threshold value of stress for unpinning the
dislocations.  (See figs. 4 and 5 in ref. \cite{Rajesh} and also for further details.)
Interestingly, for the SOC regime, {\it we find that
  most of the spatial elements are literally on the 'fold line'
corresponding to  the marginally stable state}.

  The dynamical origin of the scaling regime in our model is similar to
that   dealt by Gil and Sornette \cite{Gil} using a subcritical Hopf
bifurcation.  More recently, a dynamical analysis of
  Zhang's model of SOC has been reported as well
  \cite{cess}. However, it must be pointed out
   that in the present case, the scaling occurs at high drive values in
contrast   to many SOC type models. In this sense,
  the dynamical regimes found in our model are similar to that observed
   in thermal convection of a box of helium gas, namely,
  periodic states $\rightarrow$ chaos $\rightarrow$ power law hard
turbulence  regime \cite{Lib} as a function of Rayleigh
  number. (Actually, chaos and hard turbulence are separated by soft
turbulence.)  Lastly, to the best of our knowledge, this is
  the first model which exhibits such a crossover not just in the context
of  the PLC effect, but as a general crossover between two distinct dynamical
regimes.

  \acknowledgements
  We thank Prof. Neuh\"auser for supplying the experimental data. This work is supported by Department of
  Science and Technology, Grant No: SP/S2/K-26/98, New Delhi, India.

\begin{figure}[!h]
  \centerline{\includegraphics[height=2.5cm,width=14cm]{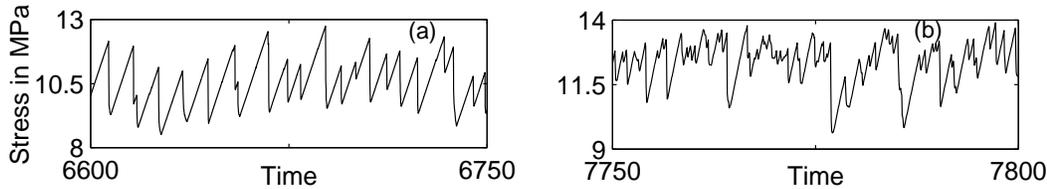}}
 \caption{Experimental stress-time series corresponding to (a) chaotic state at $\dot\epsilon_a=1.7\times 10^{-5}s^{-1}$ and (b) SOC state at
$\dot\epsilon_a=8.3\times10^{-5}s^{-1}$.}
  \end{figure}
  \begin{figure}[!h]
  \mbox{
  \includegraphics[height=4.5cm,width=6.5cm]{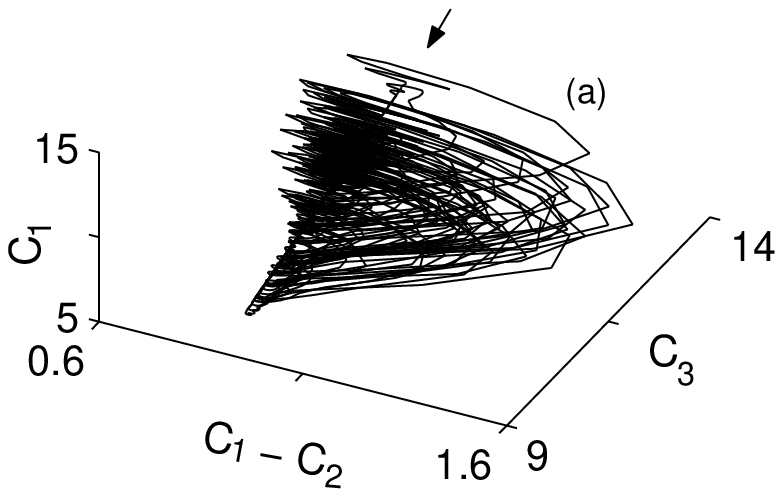}
  \hspace{0.5cm}
  \includegraphics[height=4.5cm,width=6.5cm]{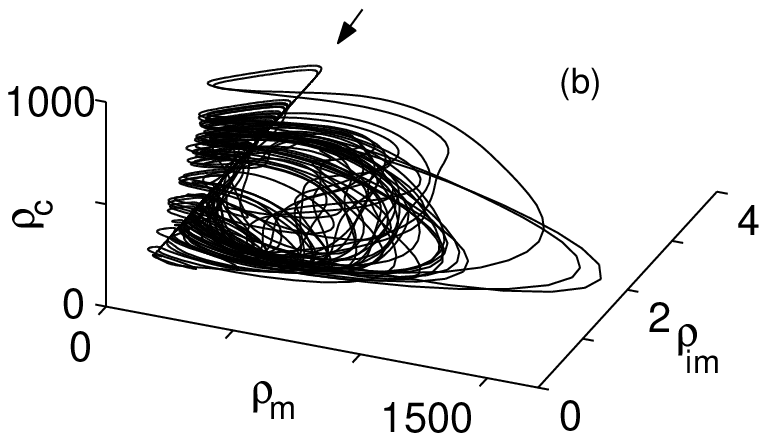}}
  \caption{(a) Reconstructed experimental attractor (b)
   Attractor from the model for $N=100$, $j=50$. }
  \end{figure}
  \begin{figure}[!h]
  \mbox{
  \includegraphics[height=3.3cm,width=6.5cm]{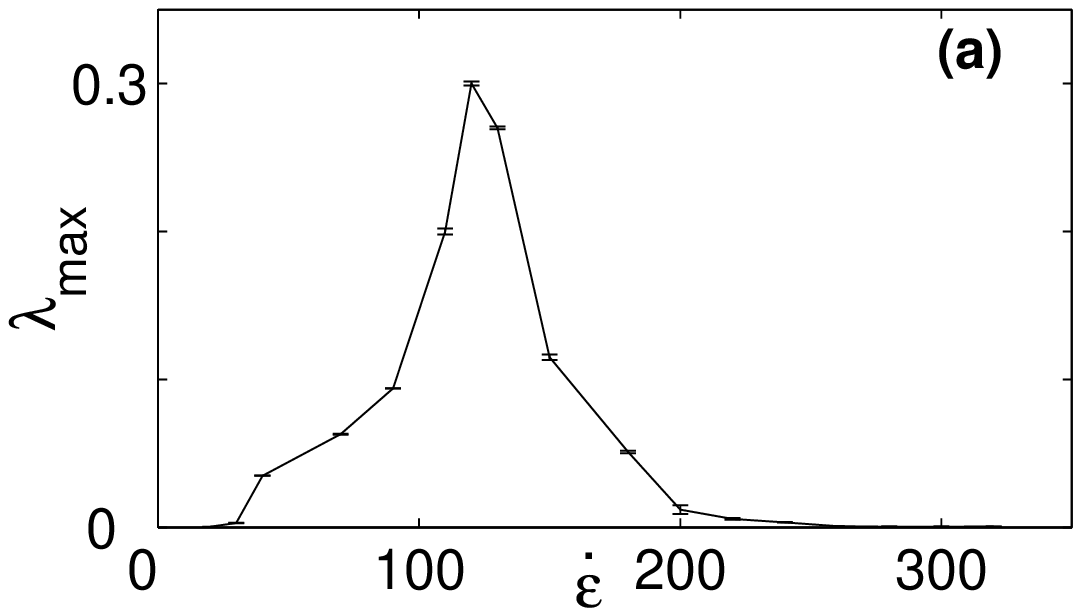}
  \hspace{0.5cm}
  \includegraphics[height=3.8cm,width=6.5cm]{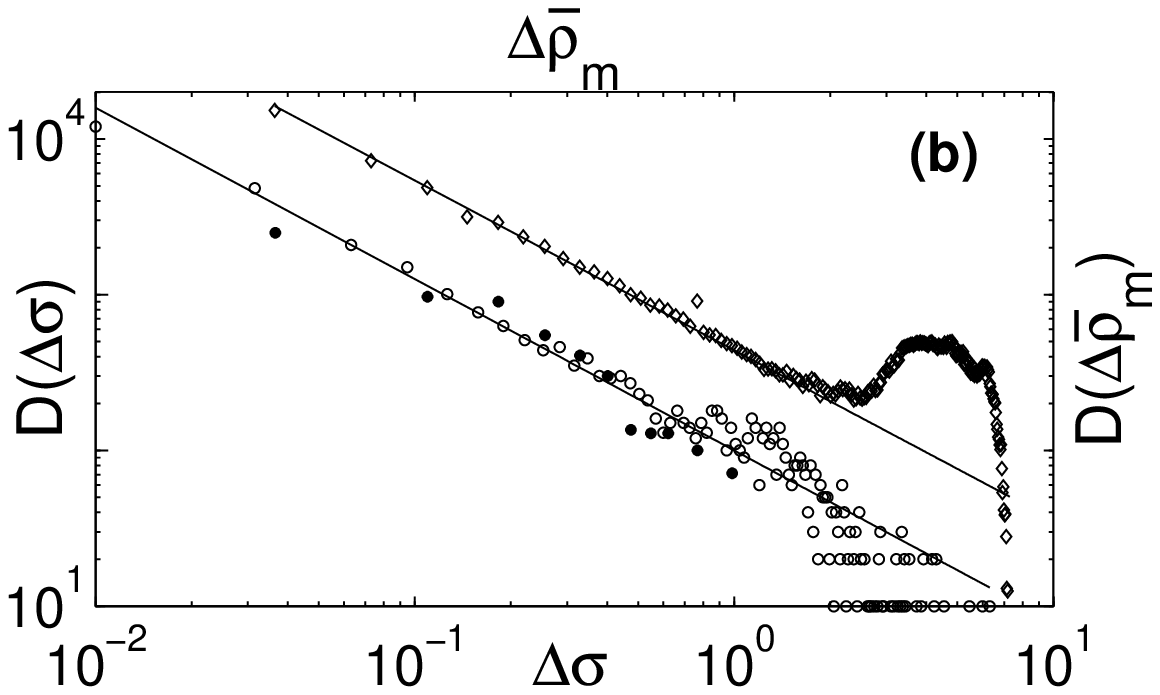}}
  \caption{(a). The largest Lyapunov exponent of the model.
  (b) Distributions of the stress drops from the model ($\circ$), from experiments
   ($\bullet$) and $\Delta\bar\rho_m$ ($\diamondsuit$) from the model. Solid    lines are guide to the eye.}
  \end{figure}

\end{document}